\documentclass[sigconf,authorversion,nonacm]{acmart}

\usepackage{dialogue}
\usepackage{epigraph}

\AtBeginDocument{%
  \providecommand\BibTeX{{%
    \normalfont B\kern-0.5em{\scshape i\kern-0.25em b}\kern-0.8em\TeX}}}

\setcopyright{cc}

\acmDOI{}
\acmISBN{}

\acmConference[HRI 2023 Workshop on Human-Robot Conversational Interaction]{HRI 2023 Workshop on Human-Robot Conversational Interaction}{March 13, 2023}{Stockholm, SE}
\begin{document}

\title{Robots Who Interrupt Talk in Meetings}

\author{Olli Niinivaara}
\email{olli.niinivaara@helsinki.fi}
\orcid{0000-0002-6846-9716}
\affiliation{%
  \institution{University of Helsinki}
  \department{Department of Practical Philosophy}
  \streetaddress{Unioninkatu 40A}
  \city{Helsinki}
  \state{}
\country{Finland}
  \postcode{FI-00014}
}

\begin{abstract}
Knowledge sharing is an important aspect in most meetings. Personal characteristics of some participants, such as their (in)ability or (un)willingness to take the floor, may  have a negative effect on the quality of knowledge sharing; some people tend to talk too much, while others have difficulties in making themselves heard. A robotic facilitator can be used to distribute the floor time more efficiently. While current research is mostly focused on encouraging participants to talk, this paper suggests interruption functionality to discourage speakers from talking. The facilitator gathers turn-taking signals from the participants and expresses them on their behalf. It hides the identity of individuals, making it easier for everyone to take action. The facilitator represents the signals coherently for all signalers, which equalizes the differences in social signalling skills, and makes it easier for the speaker to interpret the signals. It continuously gathers feedback from all participants, and thereby can represent the collective mood of the audience and smooth out outlier reactions. The facilitator can be programmed to act in a germane, courteous and attentive manner, which helps keeping the meeting mood high.

\end{abstract}

\ccsdesc[500]{Human-centered computing~Collaborative interaction}
\ccsdesc[300]{Computer systems organization~External interfaces for robotics}
\ccsdesc[100]{Social and professional topics~User characteristics}
\ccsdesc[100]{Human-centered computing~Collaborative and social computing systems and tools}

\keywords{conversational agent, automated facilitation, backchannel, interruption, turn-taking, collective feedback,
nonverbal communication}


\maketitle

\epigraph{If you had to identify, in one word, the reason why the human race has not achieved, and never will achieve, its full potential, that word would be “meetings”.}{Dave Barry}

\hypertarget{introduction}{%
\section{Introduction}\label{introduction}}

Meetings are central to organizational life, and the time people spend
in meetings is ever-increasing \citep[][]{mroz_we_2018,allen_key_2022}.
To ensure the efficiency of a meeting, one important aspect is to give
everyone an opportunity to speak out. The more participants there are,
and the less familiar the participants are with each other, the more
difficult the orchestration of turn-taking becomes. A skilled meeting
facilitator (moderator, mediator, chair, host) definitely helps, but
sometimes such a person is not available. And even then, the amount of
participants can make facilitating just too hard for humans. Therefore,
it makes sense to research various technological tools that could help
in running meetings more efficiently. In words of Woolley et
al. \citep[][]{woolley_evidence_2010}: ``Could a group's collective
intelligence be increased by, for example, better electronic
collaboration tools?''

Maybe the most sophisticated idea for orchestrating group conversations
is the use of automated facilitators
\citep[see e.g.][]{shamekhiconversational2020,bergstromvote2009,matsuyamafour-participant2015,ayllonidentification2021,skantzepredicting2017}.
However, existing systems are principally designed to distribute floor
time evenly by encouraging the more silent listeners to speak. Such an
approach has two shortcomings. One is that in order to speak, there has
to be an opportunity to do that. It is not uncommon that a small loud
and dominant minority (or even one person, say ``a besserwisser on a
mission'') can steal the show and render a meeting quite inefficient.
Therefore encouragement to speak may not be enough in itself, but a
speaker should sometimes be encouraged \emph{not} to speak, or even be
interrupted. The other shortcoming is that distributing floor time
evenly is not a goal in itself. It is namely fine that some participants
talk more than others, if they are the ones who have more important
information to communicate. It would be counterproductive to force those
participants to speak who do not happen to have anything interesting to
say.

As an addition to automated meeting facilitators, I suggest interruption
functionality that is designed specifically for discouraging the current
speaker from talking. A recent review on turn-taking in conversational
systems \citep[][]{skantze_turn-taking_2021} indicates that such a
functionality has not yet been widely considered. As exceptions,
Bergstrom \& Karahalios \citep[][]{bergstromvote2009} and Bohus \& Horvitz
\citep[][]{bohus_facilitating_2010} have considered the possibility of such
negative feedback. However, both ended up not implementing such a
functionality, and therefore do not discuss it further.

The most relevant related work is a robot for managing turn-taking by
Gonnot et al. \citep[][]{gonnot_social_2019}, which is also a system
designed to anonymously collect and then non-verbally express various
negative feedback signals. However, their input interface design
requires somewhat cognitive processing, which might be distracting. As
an alternative to that, I suggest a more minimal input interface. More
problematically, their output interface is based on robotic tractor
Cozmo\footnote{\url{https://www.digitaldreamlabs.com/pages/meet-cozmo}},
that communicates by steering around a meeting table. This communication
media is quite unnatural and obviously does not scale to larger
conferences. Instead, I suggest that a humanoid robot is used to express
the social cues, which should be much easier to interpret. And, above
all, Gonnot et al. avoid interrupting a talk, whereas here it is one of
the main objectives.

\hypertarget{input}{%
\section{Input}\label{input}}

The most important UX input requirements in this context seem to be
accuracy, user equality and discretion. Accuracy is required, because
discouraging speaking is a delicate matter, and interrupting someone is
a drastic act. An erroneous operation here would easily render a
conversational agent detrimental. User equality means that floor time
should not depend on the personal characteristics of the participants,
such as on their degree of glossophobia, extraversion, their social
skills in expressing backchanneling signals, or their abilities to do
surface acting. Discretion is required so that even the most shy and
socially anxious listeners feel safe using the system. An anonymous and
indiscernible input interface also avoids distracting the ongoing
discussion, so that the decision on whether and how to act can be
controlled by the system.

Systems that can detect the emotional intent from nonverbal
backchanneling cues are currently being developed
(\citep[e.g.][]{matsuyamafour-participant2015,ayllonidentification2021,
bohus_facilitating_2010,bousmalis_towards_2013}). However, such
an input interface would not fulfill the requirements for user equality
and discretion, because they require the users to be brave enough to
stand out, and to stand out in a way that is correctly interpreted by
the detection system. Therefore I suggest it is better to let the
participants use their own device (laptop or
smartphone\footnote{there a also Audience Response
Systems \citep[][]{kay_examining_2009} for this purpose, see
\url{https://www.rn.inf.tu-dresden.de/arselector/} for list}) to signal
their intents. The participants could, for example, surf to a web
address given in the start of meeting, and use the system by clicking on
a set of buttons (see below) offered there.

To decide the most useful set of commands, experimental work is needed.
While a lot of research on turn-taking and conversational agents exists,
research that explicitly concerns interruptions and other acts that
discourage speech is scarce. Seminal work on the area is Lycan
\citep[][]{lycan_conversation_1977}. Goldberg
\citep[][]{goldberg_interrupting_1990} gives a typology of interruptions
but focuses on vocal utterances. Schlöder \citep[][]{schloder_how_2019}
presents a taxonomy of rejection moves in dialogue. Gonnot et
al. \citep[][]{gonnot_social_2019} is an implementation with a command
palette with many useful negative feedback functionalities.

Based on the aforementioned previous work, the following action set is
suggested as a starting point. The suggested actions are divided in
three categories. Category \textbf{Advice} contains impressions
signaling that the speaker may continue, but should somehow adjust the
presentation. Category \textbf{Comment} contains signals stating that
the speaker should give the floor for a short moment, but may then
continue. Moreover, these comment signals come in two moods: stating
either that an intermediate commentary in general would be in place, or
that the signaler itself wants to utter this kind of a comment. Finally,
category \textbf{Stop} contains ways to signal the speaker to give the
floor for now.

\hspace{2cm}

Advice:

\textbf{Explain}: ``We did not understand. Please explain with more
detail''

\textbf{Doubtful}: ``We found that hard to believe. Please be more
convincing''

\textbf{Skip}: ``You are wasting our time. Please state your point on
this topic''

\hspace{2cm}

Comments:

\textbf{Questionable}: ``Let me/us ask you a question''

\textbf{Mistake}: ``Let me/us correct you''

\textbf{Dialogue}: ``Let me/us answer that''

\textbf{Announcement}: ``I have/there is a short announcement to make''

\hspace{2cm}

Stops:

\textbf{Inappropriate}: ``Your delivery does not belong here''

\textbf{Overtime}: ``Your time is up''

\textbf{Dispute}: ``You are just arguing with each other, please respect
our time and continue that somewhere else''

\textbf{Secret}: ``You cannot talk about that here''

\hspace{2cm}

In a fully working system, some auxiliary actions would probably also be
required. Some actions could be directed towards the facilitator. For
example, a \textbf{Cancel} signal could be used to cease the current
robot script when it has become irrelevant due to later happenings. Some
actions could be directed toward other audience members. For example, a
\textbf{Calm down} signal could be used to pacify a restless part of
audience.

As an extension, the signals could come in degrees. A weak signal would
mean that the signaler is in doubt, so that the output script should be
started only if some majority of listeners also entertains the same
opinion (``I wonder if there was a mistake''). A strong signal would
mean that the signaler wants the signal to be carried out in full force,
irrespective of what other opinions are currently active in the audience
(``I have an announcement to make: the building is on fire!'').

\hypertarget{output}{%
\section{Output}\label{output}}

The most important UX output requirements in this context seem to be
nondisruptiveness and effectiveness. Nondisruptiveness means that
signals should not confuse or agitate the speaker. Effectiveness means
that it should not be possible for the speaker to ignore the signals,
either deliberately or by accident.

The aforementioned requirements contradict each other. This dilemma can
be solved by using gradual, nonverbal output. Graduality means that an
output action script starts with quite inconspicuous signals like small
gestures, and progresses towards more intrusive measures, like speaking
aloud. Nonverbal output can be implement by using a humanoid
(anthropomorphic) output UI. Human gestures will be effortlessly and
immediately recognized and emotional response to them is involuntary,
without any higher cognitive processing required
\citep[][]{van_kleef_social_2022}. Urakami \& Seaborn
\citep[][]{urakami_nonverbal_2022} and Saunderson \& Nejat
\citep[][]{saunderson_how_2019} give reviews on how robots can influence
humans with such nonverbal communication.

The operation and the granularity of output is explained next with
scenarios\footnote{Note on format: Emotional responses, such as \textit{gets bored} are denoted in italics; Audience input signals ("button clicks"), such as \textbf{Mistake} are denoted in bold}.

\subsection*{scenario 1}

Let us imagine a big scientific conference where a speaker named Amy has
just made a serious mistake that should be immediately corrected.

\begin{dialogue}
\speak{Listener A} (\textit{recognizes a potential mistake}): \textbf{Mistake}
\speak{Robot} blinks eyes, slightly jerks head
\speak{More listeners} (\textit{also start to suspect a problem}): \textbf{Mistake}
\speak{Robot} squeaks eyebrows, rubs chin
\speak{More listeners} (\textit{agree that something is not right}): \textbf{Mistake}
\speak{Robot} scratches ear, shakes head, says: "hmmmm..."
\speak{Amy} (\textit{decides to react}): Says: "If you spotted something, please help me out"
\speak{Potential commentators} (\textit{willing to step up}): \textbf{Let me answer that}
\speak{Robot} raises hand, stares at the speaker
\speak{Amy} (\textit{recognizes that someone is willing to comment}): Interrupts the speech
\speak{Robot} (\textit{recognizes that speaker is silent}): Says: "We have a comment from the audience". Lowers hand. (\textit{Selects one potential commentator to get the floor})
\speak{Commentator} (\textit{Notices that robot has chosen her}): Explains the mistake
\speak{Other potential commentators}(\textit{Find the explanation adequate}): Remove their \textbf{Let me answer that} -signals
\speak{Amy} (\textit{Notes that incident is solved}) Continues the presentation
\end{dialogue}

\subsection*{scenario 2}

Let us imagine a town community public meeting (a meeting where public
can discuss with officials), where one citizen (Bob) is ranting about an
irrelevant topic.

\begin{dialogue}
\speak{Bob} keeps on talking
\speak{Listener A} (\textit{gets bored}) \textbf{Skip}
\speak{Robot} sighs, drums fingers
\speak{Bob} continues
\speak{Some more listeners} (\textit{get bored}) \textbf{Skip}
\speak{Robot} yawns and stretches, gesticulates as if looking at its wrist watch
\speak{Listener B} (\textit{gets annoyed}) \textbf{Inappropriate}
\speak{Robot} scratches forehead, shakes head, rolls eyes
\speak{Bob} starts to speak even more aggressively
\speak{Some listeners} (\textit{Hope someone to answer to Bob}) \textbf{Let's Dialogue}
\speak{Robot} raises hand, sweeps hand toward audience
\speak{Someone} (\textit{Ready to stop Bob}) \textbf{I have an Announcement}
\speak{Robot} stands up, points toward audience, coughs loudly
\speak{Bob} Keeps on talking
\speak{Majority of listeners} (\textit{Hoping Bob to shut up}) \textbf{Inappropriate} or \textbf{Skip} or \textbf{There is an announcement}
\speak{Robot} walks toward Bob, raises both hands, says: "Please, stop speaking now, or I start singing loudly."
\speak{Bob} shuts up.
\speak{Robot} signals to the announce-maker to go on, walks back to its chair and sits down
\end{dialogue}

\hypertarget{ethical-considerations}{%
\section{Ethical considerations}\label{ethical-considerations}}

Let us, for argument's sake, assume robotic facilitators to be a product
with huge sales potential. A technology that deeply disrupts how people
communicate with each other may bring benefits, but might also cause
some unintended negative consequences. Therefore it is justifiable to
consider also the potential negative effects before such robots are
being deployed in massive scale.

Maybe the most pressing issue in our times is achieving environmental
sustainability, and mass-scale production of robots would be detrimental
due to their carbon footprint and use of rare-earth elements in
actuators, batteries and motherboards. Up to a point, animated virtual
avatar facilitators could be used instead of robots. However, they might
be far less effective interruptors than physical robots. To reduce the
amount of robots, a common application programming interface for robots
from different manufacturers would enable one robot to be used in
multiple roles, meeting facilitation included. Robots could also be
shared or rented between users, for example borrowed from libraries. A
single robot could serve multiple meetings by walking from meeting to
meeting. But if we assume that a robot is already manufactured, adding
the interruption functionality to it would not cause further
environmental damage in itself, especially if the interruption component
decreases meeting times and thereby energy consumption.

Participants who do not have suitable device at hand would be outcasts
in robot-facilitated meetings. While it is realistic to assume that
everyone can bring a laptop or a smartphone, oversights and malfunctions
do happen. There are four remedies. One is to keep spare devices
available by the meeting organizer. Another is to try to automatically
detect the signals from the non-verbal (and verbal) cues of the
participants. A third is to pair people without a device with those who
have. A fourth is to not use a robot whenever someone in the audience so
wishes. As different options to handle the situation exist, missing a
device does not seem to be a critical obstacle.

An automated facilitator can record sensible data from meetings. Cloud
video conferencing software are routinely used in meetings nowadays,
hinting that electronic meeting tools in general do not pose a security
risk to organizations. However, a system that collects quantitative data
about meeting behavior poses an intraorganizational privacy risk.
Especially, using this kind of data as performance indicators for
workforce assessment might be tempting. Such data could reveal, for
example, which people agree and disagree with each other the most. Using
a facilitator robot as an intraorganizational surveillance spy should
be discouraged or technically prevented (for example by deleting all
data after every meeting). Otherwise even the awareness that every
action is being record\-ed might incentivize participants to focus their
energy on gaming the system instead of on making the meeting productive.

Continuous use of automated facilitators everywhere might lead to a
situation in which people become quite bad at communicating, cooperating
or coordinating on anything without a robot being present. It is
self-evident that advances in technology change the skills people need.
However, it is reasonable to ask, whether some basic communication
skills are so essential to us that we should not let them degenerate. On
the positive side, people are naturally quite good at imitation, and
therefore watching a robot might teach people how to facilitate well.
People should also be encouraged to hold meetings without a robot every
now and then, in order to keep old-fashioned meeting skills at some base
level. However, our understanding of what happens when robots are placed
within groups or teams for longer times is highly limited
\citep[][]{sebo_robots_2020}, therefore more research on this area is
needed.

The user interface operations suggested here are based on how people
currently orchestrate meetings. But of course also totally novel input
and output operations could be created that have no basis in the current
social reality. Such operations might turn out to be much more efficient
than current ways of managing turn-taking. After all, current signals
were never engineered, but are more or less ad hoc results of historical
(evolutionary \citep[][]{van_kleef_social_2022}, cultural
\citep[][]{li_cooperative_2001}, and societal
\citep[e.g.][]{smith-lovininterruptions1989}) developments. In our
ancient past the individuals who were the strongest and most aggressive
might have dominated meeting outcomes. Today, the most charismatic and
socially skilled people have an advantage over the socially clumsy
introverts. In the future, those who are the best using the
technologies to their advantage might dominate both the aggressive and
the charismatic ones. Such a future would be ethically more acceptable
than the status quo in the sense that it is arguably easier to learn to
use meeting tools than to change one's personality into an extrovert.
But the basic requirement is that everyone must have an equal
opportunity to learn these meeting technologies.

\hypertarget{conclusions-and-future-work}{%
\section{Conclusions and Future
Work}\label{conclusions-and-future-work}}

Automated facilitators are an advanced technology for managing
turn-taking in meetings. Current research has mostly focused on sharing
the floor-time evenly by encouraging the passive participants to engage
more. As an additional counterpart, a functionality to discourage the
speaker from talking was suggested. The facilitator discreetly gathers
reactions from the audience, and performs the collective feedback with
gradual, polite cues that become harder and harder for the speaker to
ignore. In this way, the opportunity to interrupt the speaker does not
anymore depend on the personal characteristics or the social skills of
an individual participant, making the meeting experience more equal to
everyone. The speaker does not need to be aware and understand all the
simultaneous signals from the audience, but can concentrate on the
feedback by the facilitator.

The next research step is to implement an interruption-capable
conversational agent and run experiments with it. However, implementing
only the interruptive capabilities would probably not work, because such
a robot could bring about a quite negative meeting atmosphere. To
counterbalance, some rap\-port-build\-ing backchanneling functionality
should also be implemented.

While most HRI experiments are lab experiments, field experiments seem
more adequate here. The atmosphere and dynamics of a (heated) meeting
may be hard to create synthetically. Besides, the novelty effects
\citep[][]{smedegaard2019} of introducing a robot are probably much higher in
group settings than in dyadic interaction. To let the novelty effects
wear off, a group should use an automated facilitator in their meetings
until it becomes a routine.

Anyway, designing experiments for measuring effects on large group
behaviour seems challenging. This suggests that the scarcity of
experimental research on conversational robots as members of large
groups does not necessarily stem from uselessness of such research, but
from the hardness of doing it.


\begin{acks}
This work has been funded by NordForsk.
\end{acks}


\bibliographystyle{ACM-Reference-Format}
\bibliography{interrupt_paper}


\begin{thebibliography}{23}


\ifx \showCODEN    \undefined \def \showCODEN     #1{\unskip}     \fi
\ifx \showDOI      \undefined \def \showDOI       #1{#1}\fi
\ifx \showISBNx    \undefined \def \showISBNx     #1{\unskip}     \fi
\ifx \showISBNxiii \undefined \def \showISBNxiii  #1{\unskip}     \fi
\ifx \showISSN     \undefined \def \showISSN      #1{\unskip}     \fi
\ifx \showLCCN     \undefined \def \showLCCN      #1{\unskip}     \fi
\ifx \shownote     \undefined \def \shownote      #1{#1}          \fi
\ifx \showarticletitle \undefined \def \showarticletitle #1{#1}   \fi
\ifx \showURL      \undefined \def \showURL       {\relax}        \fi
\providecommand\bibfield[2]{#2}
\providecommand\bibinfo[2]{#2}
\providecommand\natexlab[1]{#1}
\providecommand\showeprint[2][]{arXiv:#2}

\bibitem[Allen and Lehmann-Willenbrock(2022)]%
        {allen_key_2022}
\bibfield{author}{\bibinfo{person}{Joseph~A. Allen} {and} \bibinfo{person}{Nale
  Lehmann-Willenbrock}.} \bibinfo{year}{2022}\natexlab{}.
\newblock \showarticletitle{The key features of workplace meetings:
  {Conceptualizing} the why, how, and what of meetings at work}.
\newblock \bibinfo{journal}{\emph{Organizational Psychology Review}}
  (\bibinfo{date}{Sept.} \bibinfo{year}{2022}).
\newblock
\showISSN{2041-3866}
\urldef\tempurl%
\url{https://doi.org/10.1177/20413866221129231}
\showDOI{\tempurl}


\bibitem[Ayllon et~al\mbox{.}(2021)]%
        {ayllonidentification2021}
\bibfield{author}{\bibinfo{person}{David Ayllon}, \bibinfo{person}{Ting-Shuo
  Chou}, \bibinfo{person}{Adam King}, {and} \bibinfo{person}{Yang Shen}.}
  \bibinfo{year}{2021}\natexlab{}.
\newblock \showarticletitle{Identification and {Engagement} of {Passive}
  {Subjects} in {Multiparty} {Conversations} by a {Humanoid} {Robot}}. In
  \bibinfo{booktitle}{\emph{Companion of the 2021 {ACM}/{IEEE} {International}
  {Conference} on {Human}-{Robot} {Interaction}}} \emph{(\bibinfo{series}{{HRI}
  '21 {Companion}})}. \bibinfo{publisher}{Association for Computing Machinery},
  \bibinfo{address}{New York, NY, USA}, \bibinfo{pages}{535--539}.
\newblock
\showISBNx{978-1-4503-8290-8}
\urldef\tempurl%
\url{https://doi.org/10.1145/3434074.3447229}
\showDOI{\tempurl}


\bibitem[Bergstrom and Karahalios(2009)]%
        {bergstromvote2009}
\bibfield{author}{\bibinfo{person}{Tony Bergstrom} {and}
  \bibinfo{person}{Karrie Karahalios}.} \bibinfo{year}{2009}\natexlab{}.
\newblock \showarticletitle{Vote and {Be} {Heard}: {Adding} {Back}-{Channel}
  {Signals} to {Social} {Mirrors}}. In \bibinfo{booktitle}{\emph{Proceedings of
  the 12th {IFIP} {TC} 13 {International} {Conference} on {Human}-{Computer}
  {Interaction}: {Part} {I}}} \emph{(\bibinfo{series}{{INTERACT} '09})}.
  \bibinfo{publisher}{Springer-Verlag}, \bibinfo{address}{Berlin, Heidelberg},
  \bibinfo{pages}{546--559}.
\newblock
\showISBNx{978-3-642-03654-5}
\urldef\tempurl%
\url{https://doi.org/10.1007/978-3-642-03655-2_61}
\showDOI{\tempurl}


\bibitem[Bohus and Horvitz(2010)]%
        {bohus_facilitating_2010}
\bibfield{author}{\bibinfo{person}{Dan Bohus} {and} \bibinfo{person}{Eric
  Horvitz}.} \bibinfo{year}{2010}\natexlab{}.
\newblock \showarticletitle{Facilitating multiparty dialog with gaze, gesture,
  and speech}. In \bibinfo{booktitle}{\emph{International {Conference} on
  {Multimodal} {Interfaces} and the {Workshop} on {Machine} {Learning} for
  {Multimodal} {Interaction}}} \emph{(\bibinfo{series}{{ICMI}-{MLMI} '10})}.
  \bibinfo{publisher}{Association for Computing Machinery},
  \bibinfo{address}{New York, NY, USA}, \bibinfo{pages}{1--8}.
\newblock
\showISBNx{978-1-4503-0414-6}
\urldef\tempurl%
\url{https://doi.org/10.1145/1891903.1891910}
\showDOI{\tempurl}


\bibitem[Bousmalis et~al\mbox{.}(2013)]%
        {bousmalis_towards_2013}
\bibfield{author}{\bibinfo{person}{Konstantinos Bousmalis},
  \bibinfo{person}{Marc Mehu}, {and} \bibinfo{person}{Maja Pantic}.}
  \bibinfo{year}{2013}\natexlab{}.
\newblock \showarticletitle{Towards the automatic detection of spontaneous
  agreement and disagreement based on nonverbal behaviour: {A} survey of
  related cues, databases, and tools}.
\newblock \bibinfo{journal}{\emph{Image and Vision Computing}}
  \bibinfo{volume}{31}, \bibinfo{number}{2} (\bibinfo{date}{Feb.}
  \bibinfo{year}{2013}), \bibinfo{pages}{203--221}.
\newblock
\showISSN{0262-8856}
\urldef\tempurl%
\url{https://doi.org/10.1016/j.imavis.2012.07.003}
\showDOI{\tempurl}


\bibitem[Goldberg(1990)]%
        {goldberg_interrupting_1990}
\bibfield{author}{\bibinfo{person}{Julia~A. Goldberg}.}
  \bibinfo{year}{1990}\natexlab{}.
\newblock \showarticletitle{Interrupting the discourse on interruptions: {An}
  analysis in terms of relationally neutral, power- and rapport-oriented acts}.
\newblock \bibinfo{journal}{\emph{Journal of Pragmatics}} \bibinfo{volume}{14},
  \bibinfo{number}{6} (\bibinfo{date}{Dec.} \bibinfo{year}{1990}),
  \bibinfo{pages}{883--903}.
\newblock
\showISSN{0378-2166}
\urldef\tempurl%
\url{https://doi.org/10.1016/0378-2166(90)90045-F}
\showDOI{\tempurl}


\bibitem[Gonnot et~al\mbox{.}(2019)]%
        {gonnot_social_2019}
\bibfield{author}{\bibinfo{person}{Alix Gonnot}, \bibinfo{person}{Christine
  Michel}, \bibinfo{person}{Jean-Charles Marty}, {and} \bibinfo{person}{Amélie
  Cordier}.} \bibinfo{year}{2019}\natexlab{}.
\newblock \showarticletitle{Social {Robot} as an {Awareness} {Tool} to {Help}
  {Regulate} {Collaboration}}. In \bibinfo{booktitle}{\emph{Workshop on
  {Robots} for {Learning} ({R4L}) at the 28th {IEEE} {International}
  {Conference} on {Robot} \& {Human} {Interactive} {Communication}}}.
  \bibinfo{address}{New Delhi, India}.
\newblock
\urldef\tempurl%
\url{https://hal.archives-ouvertes.fr/hal-02317785}
\showURL{%
\tempurl}


\bibitem[Kay and LeSage(2009)]%
        {kay_examining_2009}
\bibfield{author}{\bibinfo{person}{Robin~H. Kay} {and} \bibinfo{person}{Ann
  LeSage}.} \bibinfo{year}{2009}\natexlab{}.
\newblock \showarticletitle{Examining the benefits and challenges of using
  audience response systems: {A} review of the literature}.
\newblock \bibinfo{journal}{\emph{Computers \& Education}}
  \bibinfo{volume}{53}, \bibinfo{number}{3} (\bibinfo{date}{Nov.}
  \bibinfo{year}{2009}), \bibinfo{pages}{819--827}.
\newblock
\showISSN{0360-1315}
\urldef\tempurl%
\url{https://doi.org/10.1016/j.compedu.2009.05.001}
\showDOI{\tempurl}


\bibitem[Li(2001)]%
        {li_cooperative_2001}
\bibfield{author}{\bibinfo{person}{Han~Z. Li}.}
  \bibinfo{year}{2001}\natexlab{}.
\newblock \showarticletitle{Cooperative and {Intrusive} {Interruptions} in
  {Inter}- and {Intracultural} {Dyadic} {Discourse}}.
\newblock \bibinfo{journal}{\emph{Journal of Language and Social Psychology}}
  \bibinfo{volume}{20}, \bibinfo{number}{3} (\bibinfo{date}{Sept.}
  \bibinfo{year}{2001}), \bibinfo{pages}{259--284}.
\newblock
\showISSN{0261-927X}
\urldef\tempurl%
\url{https://doi.org/10.1177/0261927X01020003001}
\showDOI{\tempurl}


\bibitem[Lycan(1977)]%
        {lycan_conversation_1977}
\bibfield{author}{\bibinfo{person}{William~G. Lycan}.}
  \bibinfo{year}{1977}\natexlab{}.
\newblock \showarticletitle{Conversation, politeness, and interruption}.
\newblock \bibinfo{journal}{\emph{Linguistics}} \bibinfo{volume}{10},
  \bibinfo{number}{1-2} (\bibinfo{date}{March} \bibinfo{year}{1977}),
  \bibinfo{pages}{23--53}.
\newblock
\showISSN{0031-1251}
\urldef\tempurl%
\url{https://doi.org/10.1080/08351819709370438}
\showDOI{\tempurl}


\bibitem[Matsuyama et~al\mbox{.}(2015)]%
        {matsuyamafour-participant2015}
\bibfield{author}{\bibinfo{person}{Yoichi Matsuyama}, \bibinfo{person}{Iwao
  Akiba}, \bibinfo{person}{Shinya Fujie}, {and} \bibinfo{person}{Tetsunori
  Kobayashi}.} \bibinfo{year}{2015}\natexlab{}.
\newblock \showarticletitle{Four-participant group conversation: {A}
  facilitation robot controlling engagement density as the fourth participant}.
\newblock \bibinfo{journal}{\emph{Computer Speech \& Language}}
  \bibinfo{volume}{33}, \bibinfo{number}{1} (\bibinfo{date}{Sept.}
  \bibinfo{year}{2015}), \bibinfo{pages}{1--24}.
\newblock
\showISSN{0885-2308}
\urldef\tempurl%
\url{https://doi.org/10.1016/j.csl.2014.12.001}
\showDOI{\tempurl}


\bibitem[Mroz et~al\mbox{.}(2018)]%
        {mroz_we_2018}
\bibfield{author}{\bibinfo{person}{Joseph~E. Mroz}, \bibinfo{person}{Joseph~A.
  Allen}, \bibinfo{person}{Dana~C. Verhoeven}, {and}
  \bibinfo{person}{Marissa~L. Shuffler}.} \bibinfo{year}{2018}\natexlab{}.
\newblock \showarticletitle{Do {We} {Really} {Need} {Another} {Meeting}? {The}
  {Science} of {Workplace} {Meetings}}.
\newblock \bibinfo{journal}{\emph{Current Directions in Psychological Science}}
  \bibinfo{volume}{27}, \bibinfo{number}{6} (\bibinfo{date}{Dec.}
  \bibinfo{year}{2018}), \bibinfo{pages}{484--491}.
\newblock
\showISSN{0963-7214}
\urldef\tempurl%
\url{https://doi.org/10.1177/0963721418776307}
\showDOI{\tempurl}


\bibitem[Saunderson and Nejat(2019)]%
        {saunderson_how_2019}
\bibfield{author}{\bibinfo{person}{Shane Saunderson} {and}
  \bibinfo{person}{Goldie Nejat}.} \bibinfo{year}{2019}\natexlab{}.
\newblock \showarticletitle{How {Robots} {Influence} {Humans}: {A} {Survey} of
  {Nonverbal} {Communication} in {Social} {Human}–{Robot} {Interaction}}.
\newblock \bibinfo{journal}{\emph{International Journal of Social Robotics}}
  \bibinfo{volume}{11}, \bibinfo{number}{4} (\bibinfo{date}{Aug.}
  \bibinfo{year}{2019}), \bibinfo{pages}{575--608}.
\newblock
\showISSN{1875-4805}
\urldef\tempurl%
\url{https://doi.org/10.1007/s12369-019-00523-0}
\showDOI{\tempurl}


\bibitem[Schlöder and Fernández(2019)]%
        {schloder_how_2019}
\bibfield{author}{\bibinfo{person}{Julian~J. Schlöder} {and}
  \bibinfo{person}{Raquel Fernández}.} \bibinfo{year}{2019}\natexlab{}.
\newblock \showarticletitle{How to {Reject} {What} in {Dialogue}}. In
  \bibinfo{booktitle}{\emph{Proceedings of the 23rd {Workshop} on the
  {Semantics} and {Pragmatics} of {Dialogue}}}. \bibinfo{publisher}{SemDial},
  \bibinfo{address}{London, United Kingdom}.
\newblock
\urldef\tempurl%
\url{http://semdial.org/anthology/Z19-Schlder_semdial_0014.pdf}
\showURL{%
\tempurl}


\bibitem[Sebo et~al\mbox{.}(2020)]%
        {sebo_robots_2020}
\bibfield{author}{\bibinfo{person}{Sarah Sebo}, \bibinfo{person}{Brett Stoll},
  \bibinfo{person}{Brian Scassellati}, {and} \bibinfo{person}{Malte~F. Jung}.}
  \bibinfo{year}{2020}\natexlab{}.
\newblock \showarticletitle{Robots in {Groups} and {Teams}: {A} {Literature}
  {Review}}.
\newblock \bibinfo{journal}{\emph{Proceedings of the ACM on Human-Computer
  Interaction}} \bibinfo{volume}{4}, \bibinfo{number}{CSCW2}
  (\bibinfo{date}{Oct.} \bibinfo{year}{2020}), \bibinfo{pages}{176:1--176:36}.
\newblock
\urldef\tempurl%
\url{https://doi.org/10.1145/3415247}
\showDOI{\tempurl}


\bibitem[Shamekhi(2020)]%
        {shamekhiconversational2020}
\bibfield{author}{\bibinfo{person}{Ameneh Shamekhi}.}
  \bibinfo{year}{2020}\natexlab{}.
\newblock \emph{\bibinfo{title}{Conversational {Agents} {For} {Automated}
  {Group} {Meeting} {Facilitation}}}.
\newblock Ph.{D}. {Thesis}. \bibinfo{school}{Northeastern University}.
\newblock
\urldef\tempurl%
\url{https://repository.library.northeastern.edu/files/neu:m046qf41d}
\showURL{%
\tempurl}


\bibitem[Skantze(2017)]%
        {skantzepredicting2017}
\bibfield{author}{\bibinfo{person}{Gabriel Skantze}.}
  \bibinfo{year}{2017}\natexlab{}.
\newblock \showarticletitle{Predicting and {Regulating} {Participation}
  {Equality} in {Human}-robot {Conversations}: {Effects} of {Age} and
  {Gender}}. In \bibinfo{booktitle}{\emph{Proceedings of the 2017 {ACM}/{IEEE}
  {International} {Conference} on {Human}-{Robot} {Interaction}}}
  \emph{(\bibinfo{series}{{HRI} '17})}. \bibinfo{publisher}{Association for
  Computing Machinery}, \bibinfo{address}{New York, NY, USA},
  \bibinfo{pages}{196--204}.
\newblock
\showISBNx{978-1-4503-4336-7}
\urldef\tempurl%
\url{https://doi.org/10.1145/2909824.3020210}
\showDOI{\tempurl}


\bibitem[Skantze(2021)]%
        {skantze_turn-taking_2021}
\bibfield{author}{\bibinfo{person}{Gabriel Skantze}.}
  \bibinfo{year}{2021}\natexlab{}.
\newblock \showarticletitle{Turn-taking in {Conversational} {Systems} and
  {Human}-{Robot} {Interaction}: {A} {Review}}.
\newblock \bibinfo{journal}{\emph{Computer Speech \& Language}}
  \bibinfo{volume}{67} (\bibinfo{date}{May} \bibinfo{year}{2021}),
  \bibinfo{pages}{101178}.
\newblock
\showISSN{0885-2308}
\urldef\tempurl%
\url{https://doi.org/10.1016/j.csl.2020.101178}
\showDOI{\tempurl}


\bibitem[Smedegaard(2019)]%
        {smedegaard2019}
\bibfield{author}{\bibinfo{person}{Catharina~Vesterager Smedegaard}.}
  \bibinfo{year}{2019}\natexlab{}.
\newblock \showarticletitle{Reframing the Role of Novelty within Social HRI:
  from Noise to Information}. In \bibinfo{booktitle}{\emph{2019 14th ACM/IEEE
  International Conference on Human-Robot Interaction (HRI)}}.
  \bibinfo{publisher}{IEEE}, \bibinfo{pages}{411--420}.
\newblock
\urldef\tempurl%
\url{https://doi.org/10.1109/HRI.2019.8673219}
\showDOI{\tempurl}


\bibitem[Smith-Lovin and Brody(1989)]%
        {smith-lovininterruptions1989}
\bibfield{author}{\bibinfo{person}{Lynn Smith-Lovin} {and}
  \bibinfo{person}{Charles Brody}.} \bibinfo{year}{1989}\natexlab{}.
\newblock \showarticletitle{Interruptions in {Group} {Discussions}: {The}
  {Effects} of {Gender} and {Group} {Composition}}.
\newblock \bibinfo{journal}{\emph{American Sociological Review}}
  \bibinfo{volume}{54}, \bibinfo{number}{3} (\bibinfo{year}{1989}),
  \bibinfo{pages}{424--435}.
\newblock
\showISSN{0003-1224}
\urldef\tempurl%
\url{https://doi.org/10.2307/2095614}
\showDOI{\tempurl}


\bibitem[Urakami and Seaborn(2022)]%
        {urakami_nonverbal_2022}
\bibfield{author}{\bibinfo{person}{Jacqueline Urakami} {and}
  \bibinfo{person}{Katie Seaborn}.} \bibinfo{year}{2022}\natexlab{}.
\newblock \showarticletitle{Nonverbal {Cues} in {Human}-robot {Interaction}:
  {A} {Communication} {Studies} {Perspective}}.
\newblock \bibinfo{journal}{\emph{ACM Transactions on Human-Robot Interaction}}
  (\bibinfo{date}{Dec.} \bibinfo{year}{2022}).
\newblock
\urldef\tempurl%
\url{https://doi.org/10.1145/3570169}
\showDOI{\tempurl}


\bibitem[van Kleef and Côté(2022)]%
        {van_kleef_social_2022}
\bibfield{author}{\bibinfo{person}{Gerben~A. van Kleef} {and}
  \bibinfo{person}{Stéphane Côté}.} \bibinfo{year}{2022}\natexlab{}.
\newblock \showarticletitle{The {Social} {Effects} of {Emotions}}.
\newblock \bibinfo{journal}{\emph{Annual Review of Psychology}}
  \bibinfo{volume}{73}, \bibinfo{number}{1} (\bibinfo{year}{2022}),
  \bibinfo{pages}{629--658}.
\newblock
\urldef\tempurl%
\url{https://doi.org/10.1146/annurev-psych-020821-010855}
\showDOI{\tempurl}


\bibitem[Woolley et~al\mbox{.}(2010)]%
        {woolley_evidence_2010}
\bibfield{author}{\bibinfo{person}{Anita~Williams Woolley},
  \bibinfo{person}{Christopher~F. Chabris}, \bibinfo{person}{Alex Pentland},
  \bibinfo{person}{Nada Hashmi}, {and} \bibinfo{person}{Thomas~W. Malone}.}
  \bibinfo{year}{2010}\natexlab{}.
\newblock \showarticletitle{Evidence for a {Collective} {Intelligence} {Factor}
  in the {Performance} of {Human} {Groups}}.
\newblock \bibinfo{journal}{\emph{Science}} \bibinfo{volume}{330},
  \bibinfo{number}{6004} (\bibinfo{date}{Oct.} \bibinfo{year}{2010}),
  \bibinfo{pages}{686--688}.
\newblock
\urldef\tempurl%
\url{https://doi.org/10.1126/science.1193147}
\showDOI{\tempurl}


\end{thebibliography}

\end{document}